\documentclass{ws-acs}

\usepackage{xcolor}

\begin{document}

\makeatletter
\def\@biblabel#1{[#1]}
\makeatother

\markboth{J. Falk et al.}{Structural insulators and promotors in networks under generic problem-solving dynamics}

%
\catchline{}{}{}{}{}
%

\title{STRUCTURAL INSULATORS AND PROMOTORS IN NETWORKS UNDER GENERIC PROBLEM-SOLVING DYNAMICS}

\author{JOHANNES FALK$^\dagger$\footnote{Corresponding Author}~, EDWIN EICHLER$^{\ddagger \mathsection}$, KATJA WINDT$^{\ddagger \mathparagraph}$ and MARC-THORSTEN~HÜTT$^\dagger$}
\address{
$^\dagger$ School of Science,\\
Constructor University,\\
Campus Ring 1, 28759 Bremen,\\
Germany\\[3pt]
$^\ddagger$ SMS Group GmbH,\\
40237 Düsseldorf,\\
Germany\\[3pt]
$^\mathsection$ EICHLER Consulting AG,\\
6353 Weggis,\\
Switzerland\\[3pt]
$^\mathparagraph$ School of Business, Social and Decision Sciences,\\
Constructor University,\\
Campus Ring 1, 28759 Bremen,\\
Germany\\[3pt]
$^*$jfalk@constructor.university
}

\maketitle

\begin{history}
\received{(received date)}
\revised{(revised date)}
\end{history}

\begin{abstract} 
The collective coordination of distributed tasks in a complex system can be represented as decision dynamics on a graph. This abstract representation allows studying the performance of local decision heuristics as a function of task complexity and network architecture. Here we identify hard-to-solve and easy-to-solve networks in a social differentiation task within the basic model of small-world graphs. 
We show that, depending on the details of the decision heuristic as well as the length of the added links, shortcuts can serve as \textit{structural promotors}, which speed up convergence towards a solution, but also as \textit{structural insulators}, which make the network more difficult to solve. 
Our findings have implications for situations where, in distributed decision systems, regional solutions emerge, which are globally incompatible as for example during the emergence of technological standards.
\end{abstract}

\keywords{Graph Coloring Dynamics, Distributed Decision Strategies, Global Coordination, Self-organized dynamics}

\section{Introduction}
\label{sec:intro}
Self-organized dynamics on graphs are an important concept to analyze distributed decision-making and task coordination. Beyond social sciences~\cite{kearnsExperimentalStudyColoring2006,kearnsBehavioralExperimentsBiased2009b,shiradoLocallyNoisyAutonomous2017} also logistics~\cite{grolikInformationLogisticsDecentralized2012} and computer science~\cite{buttAgentBasedModelingDistributed2020,hernandezDistributedGraphColoring2011,leithSelfManagedDistributedChannel2006a} are interested in how distributed decisions can efficiently lead to global coordination, e.g., to avoid queuing or to minimize interference between wireless networks. 
In the simplest coordination problems, a node of the graph can select a decision (a ’color’)
out of a list of allowed decisions based on the observed
decision states of its direct neighbors.
The local decision heuristics (i.e., the decision selection criteria at each node) represent the goal of the systemic task. Such coordination tasks come in two variants~\cite{juddBehavioralDynamicsInfluence2010b}: Either the task is related to some
type of consensus across the whole system. In this case,
the graph is ’solved’, when no different colors are linked.
Alternatively, these coordination tasks can be related to
social differentiation, scheduling, or resource allocation.
In this case, the graph is ’solved’, when no same colors are
linked. Here we focus on the second scenario of social differentiation and scheduling. Its abstraction as color dynamics on graphs, related to the graph coloring problem, has been made popular by the seminal work of Kearns
et al.~\cite{kearnsExperimentalStudyColoring2006}. This framework has led to relevant insight
into problem-solving dynamics and some ’stylized facts’
about distributed decision making. Examples include the positive effect of random agents in a distributed decision
system~\cite{shiradoLocallyNoisyAutonomous2017}, the effect of a wave-like organization of attention and strategic waiting on these decision dynamics~\cite{hadzhievMODELGRAPHCOLORING2009},
and the effect of shortcuts in a small-world architecture on the convergence toward a fully solved system. 
This is visible, both in experiments
with human subjects~\cite{kearnsExperimentalStudyColoring2006} and numerical simulations involving simple heuristics~\cite{hadzhievMODELGRAPHCOLORING2009}.

The decision heuristics introduced in Hadzhiev et al.~\cite{hadzhievMODELGRAPHCOLORING2009} furthermore provided a better understanding of the interplay
of centralized and autonomous, decentralized control in
manufacturing planning and control~\cite{windtGraphColoringDynamics2010,blunckBalanceAutonomousCentralized2018}.

However, a striking characteristic of graph coloring dynamics has not been understood in the past: For a fixed number of a few shortcuts (i.e., for example, as a result of a small rewiring probability in the Watts-Strogatz model [26]) one observes a dramatic variability of runtime.
Here we show that – besides the random initialization, as well as the general stochastic nature of these dynamics – this high variability is due to the network topology: Depending on the exact positions as well as the heuristic employed, shortcuts in a ring graph can generate easy-to-solve and difficult-to-solve graphs. They can act as \textit{structural insulators} or \textit{structural promotors}, i.e., they either delay or accelerate regional reorganization efforts towards a trans-regionally compatible solution.

The problem we address is of relevance for many real-world applications: In these dynamics, regional solutions emerge
rapidly, but they are incompatible on a global scale and the diffusing remaining conflicts, which are the boundaries of incompatible solution regimes, require an excessive amount of local reorganization, until one region switches to a solution compatible with another region. This problem of different locally valid solutions that are globally incompatible can especially be observed in the emergence of compatibility standards~\cite{vandekaaEmergenceStandardsMetaAnalysis2007}: Different technical devices may be locally compatible based on one standard, but incompatible with functionally equivalent standards from other areas, leading to competition between alternatives~\cite{suarezDominantDesignsSurvival1995} and ultimately resulting in a global standard. Examples of such battles are BlueRay vs HD DVD or Wi-Fi vs HomeRF~\cite{vandekaaFactorsWinningFormat2015}. There already exist some models to explain the success or failure of standards. But as economic models, they are focused on the interplay of strategic factors, business models, and business actors~\cite{papachristosSystemDynamicsModel2021,caseyDynamicsTwosidedPlatform2012}. Our investigation rather contributes to understanding the spatial organization of standards and hence the influence of the network topology on the time until a standard settles.

\section{Methods}
\label{sec:methods}
We investigate heuristics that can solve the graph coloring problem based on local decisions. In this problem from \textit{graph theory}, the goal is to assign colors to the vertices of a graph such that no two adjacent vertices have the same color. The minimum number of colors that are needed to color a network in this way is known as the \textit{chromatic number} $\chi$ of the graph. 
In this section, we explain how we generate graphs with a given chromatic number, introduce different local decision heuristics, and present a genetic algorithm that we use to generate networks with specific properties.

\subsection{Small-World Networks}

\label{sec:small_world}
In this analysis, we mainly focus on small-world networks with few inserted links as a toy model for graphs with high clustering and small shortest path length. The idea of the graph generation follows~\cite{wattsCollectiveDynamicsSmallworld1998}. However, since the networks are supposed to be solvable with a given number of $\chi$ colors (the chromatic number), we generate them as follows:
40 (39 for $\chi=3$) nodes are arranged as a circular graph, where each node $i$ is connected to its $\chi - 1$ closest neighbors in both directions. A given number of shortcuts are then added such that each shortcut connects only nodes with a different value of $mod(i, \chi)$, where $i$ is the node index, thus preserving the graph's chromatic number $\chi$. To compare how fast different network topologies can be solved, we look at the number of color changes that have been performed until the network is in a solved state. The color changes then set a time scale where each time step is equal to one color change. 

\subsection{Other graph topologies with $\chi = 2$}
\label{sec:graph_generator}
To extend our results to more general statements we generate three other types of random networks (only for $\chi = 2$):
\begin{itemize}
\item \textbf{BA}: For this network, we start with a simple path graph with 4 numbered nodes. We then add nodes and links following preferential attachment as described in~\cite{barabasiEmergenceScalingRandom1999} where each new node (labeled with a consecutive number) is attached to existing nodes via two links. However, and in contrast to the reference, to ensure that the graph has a chromatic number of $2$, for an even (odd) number of already existing nodes, a newly added node can only connect to nodes with an odd (even) label. 

\item \textbf{Random}: The procedure to create this graph starts with a graph of $N$ unconnected nodes, labeled with an integer $i$. A given number of edges is then sampled randomly from all edges that would connect two nodes with an even and an odd label. This ensures a chromatic number of $\chi = 2$. If the resulting graph is not connected, the procedure is repeated with a different set of randomly selected edges. 

\item \textbf{Modular (Mx)}: To generate this graph, we start with two separate graphs $A$ and $B$ of type \textit{random}. We then rewire $x$ randomly selected edges so that each edge connects one node from $A$ and one from $B$. Similar to the procedure for small-world networks, the connections are always added in such a way that the chromatic number $\chi = 2$ is preserved. For small $x$ the graph has high modularity. The larger $x$ the, the more similar the graph becomes to a random graph.
\end{itemize}

\subsection{Neighborhood assessment strategies}
\label{sec:heuristics}

Agent-based models to solve graph coloring problems have already been analyzed in various variations. Inspired by the results from~\cite{kearnsExperimentalStudyColoring2006}, Hadzhiev et al.~\cite{hadzhievMODELGRAPHCOLORING2009} developed a family of local decision heuristics that allow agent-based networks to be solved in reasonably short times.
Following the concepts from~\cite{hadzhievMODELGRAPHCOLORING2009}, a graph coloring heuristic consists of two components: One strategy for the temporal organization (indicating which node acts next) and one for the neighborhood assessment (indicating which color the active node selects). To simulate the behavior of independent distributed systems as closely as possible, we always use random sequential updates (R) for the temporal organization, which means that every time step the next node is selected at random from all available nodes. Using other heuristics for temporal organization, e.g. the channeled attention strategy (C) from~\cite{hadzhievMODELGRAPHCOLORING2009}, the results are qualitatively similar (data not shown).
For the neighborhood assessment heuristic, we first refer to three strategies from~\cite{hadzhievMODELGRAPHCOLORING2009}, namely $R$ (random), $M$ (color minimizing), and $W$ (strategic waiting). We then present a new ($N$) heuristic whose behavior can be continuously tuned by a parameter $r$ (reasoning): For large values of $r$ the agents always select their color by reasoned considerations. The smaller $r$, the more often the color choice happens randomly.
In all strategies, the active node first assesses the colors of its connected neighbors. If possible, the node randomly selects one of the colors that does not appear in its neighborhood (conflict-free color). Otherwise, the different strategies proceed as follows:
\begin{itemize}
    \item \textbf{R (random color):} The node selects a color at random from all available colors
    \item \textbf{M (conflict minimizing color):} The node selects randomly a color from the set of colors that minimizes the number of conflicts. If the node has already the unique conflict-minimizing color, a color is selected at random.
    \item \textbf{W (strategic waiting):} Equal to the M scheme, however, if the node has already the unique conflict-minimizing color, the present color is retained with probability $p = 0.9$.  
    \item \textbf{N (reasoning):} With a probability $r$ the node randomly selects a color that minimizes the conflicts (reasoned acting). In the other case (with a probability $1-r$) it randomly selects a color from the list of all available colors.
\end{itemize}
The $N$ heuristic can hence be understood as a generalization of the three other heuristics. For small $r$ the $N$ heuristic is similar to the $R$ heuristic, for intermediate $r$ it is similar to the $M$, and for large $r$ to the $W$ heuristic. 

In order to name the full heuristics, we follow the naming scheme that was also used in~\cite{hadzhievMODELGRAPHCOLORING2009}: $XY$ means that we used $X$ as temporal organization strategy and $Y$ as neighborhood assessment strategy.

\subsection{Genetic Algorithm}
\label{sec:genetic}
To assess how strongly the topology of a network (with a fixed number of shortcuts) affects the runtime, we use a genetic algorithm that evolves to easy-to-solve or hard-to-solve networks (with respect to a given heuristic). The algorithm starts with an ensemble of six randomly selected small-world networks with the given number $S$ of shortcuts and proceeds as follows:
\begin{itemize}
    \item Each network of the ensemble is randomly colored and then solved by the respective strategy. The time until solved (measured in activation steps) is averaged over 500 runs.
    \item The two fastest (slowest) solved networks are kept for the next run, additionally, four networks are generated by mutations (rewiring of one shortcut) and by recombination (take $n$ shortcuts from one network and $S-n$ shortcuts from the other network) of these two fastest (slowest) networks.
    \item These six new networks are the new ensemble for the first step.
\end{itemize}
After around 200 evolution steps, we could -- for the network sizes used in this investigation -- no longer observe any significant improvements. However, to ensure that rare modifications are also covered by the algorithm, we terminated the process after 1000 evolution steps and saved the obtained topologies.


\section{Results}
\label{sec:results}
We take the observed high variability of the distributed graph coloring problem as an opportunity to examine how the network topology influences the runtime. To focus the analysis we limit ourselves to networks with a chromatic number of $\chi = 2$. In the last part of the results section, we explain why networks with $\chi > 2$ show a significantly more complicated behavior, which results from the interaction of different mechanisms and thus defies a simple mechanistic explanation.

\begin{figure}
    \centering
    \includegraphics[width=.95\linewidth]{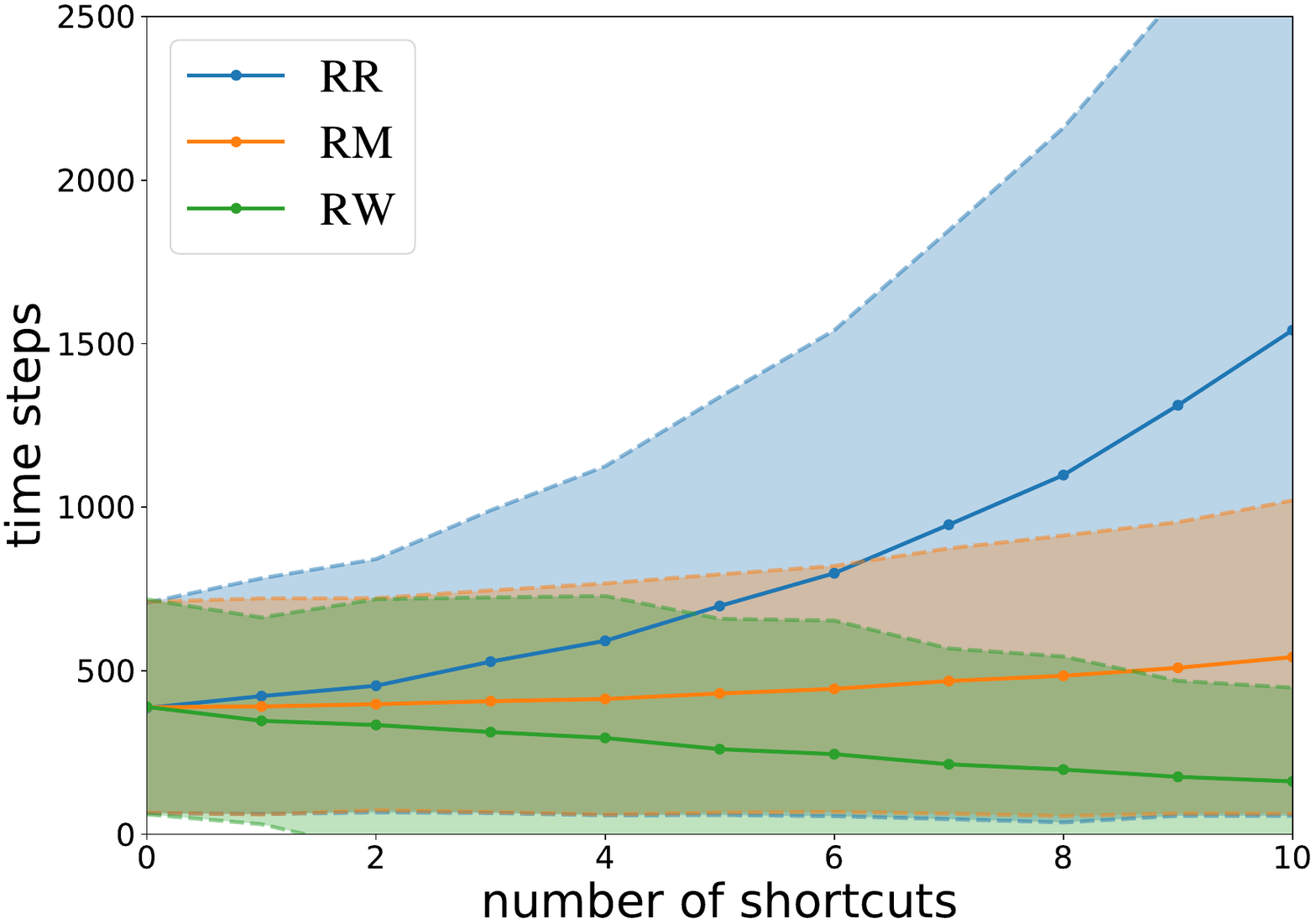}
    \caption{ Mean number of time steps (color changes) until the network is solved vs. the number of shortcuts for small-world networks using the $RR$ (random attention -- random color), $RM$ (random attention -- conflict-minimizing color), and $RW$ (random attention -- strategic waiting) heuristic. The light area denotes the standard deviation (reproduced from~\cite{hadzhievMODELGRAPHCOLORING2009}).}
    \label{fig:compare_heuristics_shortcuts}
\end{figure}

We begin our investigation by looking at some results from~\cite{hadzhievMODELGRAPHCOLORING2009}. The authors analyzed how different graph coloring heuristics perform in small-world networks when the number of shortcuts increases. In Fig.~\ref{fig:compare_heuristics_shortcuts} we show the performance of the three heuristics that use random sequential updates ($R$) for the temporal organization, and $R$, $M$ or $W$ as neighborhood assessment (see~\ref{sec:heuristics} for details). With the $RR$ and $RM$ heuristic, the more shortcuts the network has, the longer (on average) the nodes need to organize and finally solve the network. In contrast, using the $RW$ heuristic the solution is reached faster with more added links, as it was also observed in human subject networks~\cite{kearnsExperimentalStudyColoring2006}. Looking at Fig.~\ref{fig:compare_heuristics_shortcuts}, it is also noticeable that -- for a fixed number of shortcuts -- the variance of the time steps required is strikingly high. Since the initial conditions for each run are chosen randomly and the heuristic contains stochastic components, a certain variance is to be expected. An open question, however, is whether the topology, i.e. the location of the shortcuts, has an impact on the solvability.

\begin{figure}
    \centering
    \includegraphics[width=.95\linewidth]{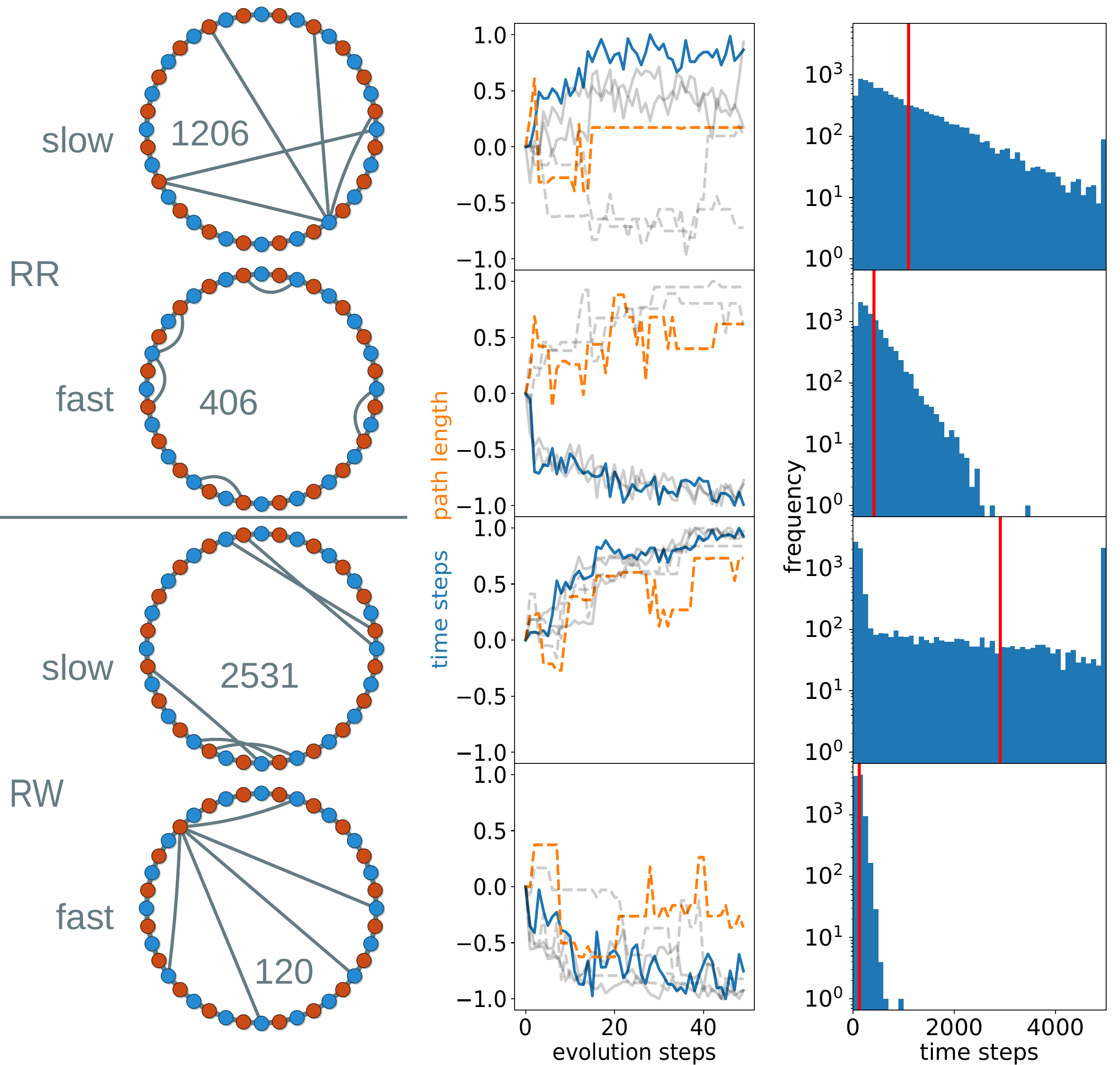}
    \caption{Comparison of the results for the genetic algorithm evolved with the $RR$ or $RW$ heuristics for hard-to-solve (slow) and easy-to-solve (fast) networks with 5 shortcuts. (first column) Exemplary topology after 50 evolution steps. The numbers indicate how many time steps the heuristics required to solve the respective graph (averaged over 500 random initial conditions). (second column) Required time steps (solid, blue) and the average path length (dashed, orange) for the networks vs. the evolution steps. The solid grey and dashed grey lines show the trajectories of other runs. The data was scaled to a range of [-1,1]. (third column) histogram of the time steps require to solve an evolved network with the given heuristic (for 10,000 random initial conditions). The last bin contains all data above the upper limit of 5,000 time steps. The red line indicates the mean.}
    \label{fig:genetic_alg}
\end{figure}

To test and quantify the impact of the topology, we use a genetic algorithm (see~\ref{sec:genetic}) that is designed to generate easy and hard-to-solve small-world graphs with a small number of 5 added links. A strong difference between the runtimes of the extreme graphs could indicate whether and how the topology affects the runtime. Results of the network evolution for the $RR$, as well as the $RW$ heuristic, are presented in Fig.~\ref{fig:genetic_alg}. The large difference between the fastest and slowest networks ($120$ vs. $2531$ color changes for $RW$ heuristic, $406$ vs. $1206$ color changes for the $RR$ heuristic) indicates that -- for a fixed number of shortcuts -- the runtimes depend strongly on the shortcut positions. Additionally, the resulting topologies seem to have characteristic features (see also second column of Fig.\ref{fig:genetic_alg}): Long-range links facilitate a fast solution finding for the $RW$ heuristic, but create a difficult-to-solve network for the $RR$ heuristic. Likewise, the easy-to-solve network for the $RR$ heuristic is characterized by maximally short links, whereas for the $RW$ heuristic the short links appear in the difficult graph.

In what follows we will introduce a generalized heuristic and extract general features that can explain the interdependence between topology and runtime. Long-range links are often considered to be beneficial for a system-wide organization because they allow transmitting information over a long distance~\cite{araujoLongrangeConnectionsRealworld2022}. Our analysis is based on the idea that the respective agent must be able to process the additional information provided by a long link. When agents evaluate the observations from their neighborhood reasoned, the remote information helps them to adapt themselves to the global solution. If, on the other hand, the agents do not operate reasoned, the additional source of information creates confusion, which hinders the stabilization of local solutions. To test this proposition, we introduce a new heuristic $N$. This heuristic can be continuously adjusted between reasoned and random behavior by means of a single parameter $r$ (details in Sec.~\ref{sec:heuristics}).

\begin{figure}
    \centering
    \includegraphics[width=.95\linewidth]{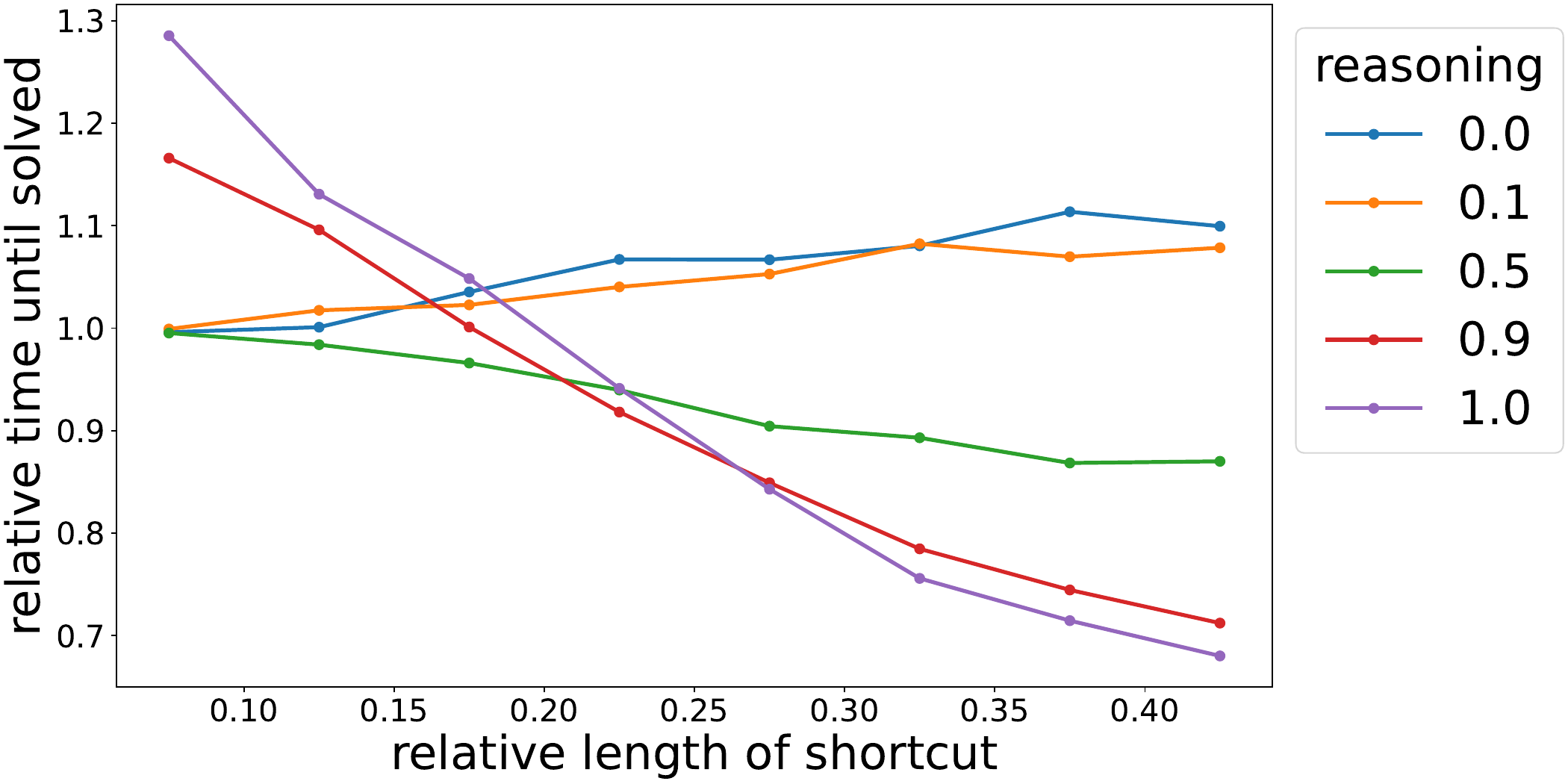}
    \caption{Relative extra time until ring graphs with 40 nodes and a single shortcut are
solved vs. the relative length of the added shortcut for different values of $r$. A relative length of $1$ refers to a shortcut of maximal length, hence spanning 20 nodes. The time is measured relative to the time that is needed if the ring graph does not have any shortcut.}
    \label{fig:rational}
\end{figure}

We create a ring lattice with 40 nodes and add a single shortcut (with the constraint that the chromatic number $\chi = 2$ is conserved, see also Sec.~\ref{sec:small_world}) For Fig.~\ref{fig:rational} we set $r$ to different values and analyze how the runtime depends on the relative length of the added shortcut (averaged over 10.000 runs each). As expected, if the heuristic is very reasoned (large $r$) the time until solved decreases for longer shortcuts. In contrast, if the heuristic contains a lot of randomnesses (small $r$), long-range links deteriorate the solvability of the graph. An additional observation is that the reasoned strategies work poorly, when the inserted link is very short (an increase of the required time by about 30\%). 

\subsection{Reasoned Agents (large $r$)}
\begin{figure}
    \centering
    \includegraphics[width=.95\linewidth]{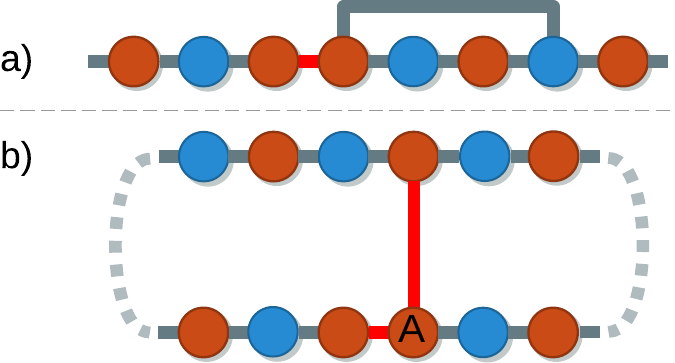}
    \caption{Comparison of the two effects a shortcut can have: (a) A short link stabilizes a solution regime against perturbations from the outside. In the example, there is a color conflict between the two red nodes (indicated by a red link). The right red node has two blue neighbors (one direct and one via the shortcut). If the node acts reasoned its color is stabilized since red minimizes the conflicts. (b) The sketch shows two sections of a large ring graph (indicated by the gray dashed line). The long shortcut organizes two distant sections and orders them. Without the shortcut, the node with the label ``A'' would have a 50\% chance of keeping its color, compared to changing to blue. Due to the shortcut, reasoned-acting nodes will change to blue, since this is the conflict-minimizing color. }
    \label{fig:gcd_conflict_examples}
\end{figure}

For large $r$ the results are in line with the slow network obtained for the $RW$ heuristic in Fig.~\ref{fig:genetic_alg}. The slow network is characterized by comparably short links that create two densely connected areas. These clusters foster a fast emergence of local solutions. Additionally, the short shortcuts stabilize the local solution against fluctuations from the outside. Figure~\ref{fig:gcd_conflict_examples}a shows an example of such stabilization of a local solution. The larger the parameter $r$, the more stable the locally solved areas. However, in the likely case that the local solution is not compatible with the currently prevailing global solution domain, the system is in a hard-to-solve state: The reasoned agents cling to their local solution, the added link acts as a \textit{structural insulator}. Contrarily, evolving towards topologies that are easy to solve for the $RW$ heuristic, the resulting network is characterized by a few nodes that are connected to various areas of the network and that act as \textit{ordering nodes}. These ordering nodes synchronize the local solutions already during their build-up. An example of the effect of a single long-range shortcut is shown in Figure~\ref{fig:gcd_conflict_examples}b. Without the shortcut, the node labeled with ``A'' could either stay red or change its color to blue. In both cases, the result would be a single conflict with one neighbor. However, due to the shortcut -- that is by definition inserted such that it does not alter the graph's chromatic number and, hence, a global solution is possible -- a change to blue minimizes the local color conflicts and acts as a local reference for the global solution domain.

\subsection{Irrational Agents (small $r$)}
The situation is different for irrational agents, i.e. with small $r$ (similar to the $RR$ heuristic). Here, Fig.~\ref{fig:compare_heuristics_shortcuts} tells us that shortcuts consistently create graphs that are more difficult to solve than the pure ring graph, where the effect is stronger the longer the added link. Consequently, the results from Fig.~\ref{fig:genetic_alg} show that the fast networks are characterized by short links. For the $RR$ heuristic, the difficult-to-solve networks are characterized by long-range links, very similar to the graphs that are easy to solve for the $RW$ heuristic. For irrational agents (as in the $RR$ heuristic), the long links that connect a single node to various areas of the graph act like a source of noise: A color-fluctuation of the highly connected node immediately destabilizes the colors of all connected nodes, spread over the full network. 

\subsection{Complex Topologies}

\begin{figure}
    \centering
    \includegraphics[width=.95\linewidth]{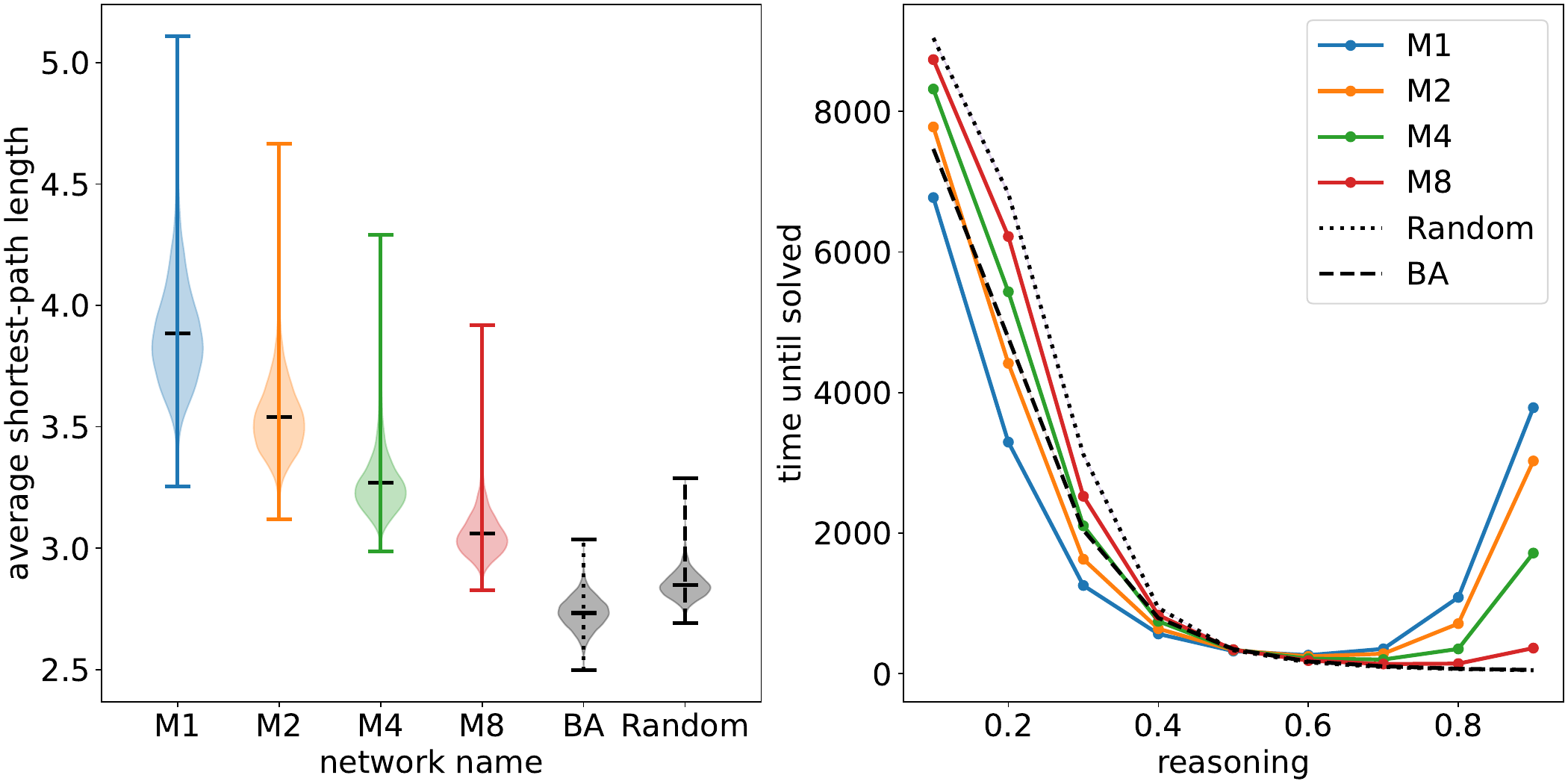}
    \caption{(left) Distributions of the average shortest-path length for the different random graphs shown in the right figure. The abbreviation $Mx$ denotes a modular graph with $x$ rewires. Each distribution contains 10,000 data-points. (right) Mean number of time steps (color changes) until the network is solved vs. the reasoning of the heuristic for different graph topologies (see Sec.~\ref{sec:graph_generator}), averaged over 10,000 networks. The standard-deviation of the mean is smaller than the markers.}
    \label{fig:random_nets}
\end{figure}

Having analyzed the interplay between the length of added links and the reasoning of the acting agents in small-world graphs, it is now natural to ask, whether this behavior can also be observed in more complex networks. As described in Sec.~\ref{sec:graph_generator}, we generated modular graphs (2 x 20 nodes, 40 edges each) with different numbers of rewires, random graphs (40 nodes, 80 edges), and BA graphs (40 nodes). All graphs are generated such that $\chi = 2$. In Fig.~\ref{fig:random_nets}(left) we show the distribution of the average shortest-path length for the different networks. For the modular graph, the more rewires we do, the shorter gets the path length. In Fig.~\ref{fig:random_nets}(right) we show the time until solved vs the reasoning parameter $r$ of the $N$ heuristic (averaged over 10,000 networks each). For both the random networks and the BA graphs, the more reasoned the agents act, the faster they are. Note, however, that for $r=1.0$ dead-lock situations are possible that cannot be solved (see e.g. Fig.~2 in~\cite{hadzhievMODELGRAPHCOLORING2009}). The results confirm the observations from the small-world networks: Random networks as well as BA networks have small modularity and high connectivity. It is therefore unlikely that globally incompatible solutions can stabilize against the rest of the network. The modular network is, however, specifically designed to have two almost separate modules. Fig.~\ref{fig:random_nets} shows that in this case heuristics that act too reasoned have a disadvantage: If the two modules converge to different solution domains, it is difficult for the heuristic to overturn one solution. The more edges we rewire, the less modular the network is. Consequently, we observe that reasoned heuristics become more advantageous with the number or rewires.   

\subsection{Extension to $\chi = 3$}

The natural extension of our investigation is to increase the chromatic number of the graphs. For Fig.~\ref{fig:chromati3_shortcuts} we performed a similar analysis as for Fig.~\ref{fig:rational}, but with a ring graph with 39 nodes and a chromatic number of $\chi = 3$. Depending on the length of the added shortcut the system takes longer or is faster to solve than without a shortcut. The general behavior of the network is on average similar to the one with a chromatic number of two (short shortcuts lead to longer times). However, there are also two drastic differences:
(1) The curve shows an alternating behavior that was not present for the $\chi = 2$ graphs. The reason is a complicated interplay between the shortcuts and the different possible solution regimes. For two colors there are only two possible solution domains: $abab$ or $baba$. However, for three colors there are $3! = 6$ possible solution domains that are facilitated or suppressed depending on the position of the shortcut. (2) The relative effect of a single shortcut is not as strong as for the $\chi = 2$ graph. The main reason is that a shortcut at each end excludes only one color at a time. If there are only two colors a single disallowed color directly determines the correct color: $\neg \text{red} \rightarrow \text{blue}$. However, the more colors we have the less effect has the banning of a single color. To control such a setting one would need to generalize the definition of a shortcut. For $\chi = 3$ such a generalized shortcut would hence consist of four conventional shortcuts that all-to-all connect two adjacent nodes with two other adjacent nodes.

\section{Conclusion}
In small-world networks, shortcuts reduce the average path length and facilitate the transport of local information through the system~\cite{marrSimilarImpactTopological2006}. One would therefore expect that distributed coordination problems on graphs always benefit from shortcuts, albeit the effect size might depend on the respective length of the shortcut. Here, we discussed the graph coloring problem as a simple form of distributed coordination problem. We analyzed how shortcuts affect the time a local heuristic needs to solve the coloring problem. Depending on how reasoned the agents act, added shortcuts give rise to different mechanisms: They synchronize the solution domains between distant sections of the network, stabilize parts of the network against fluctuations, or they create perturbations. For reasoned heuristics, short shortcuts tend to insulate locally solved but globally incompatible solutions against each other, finally leading to an increase in the overall time until a solution is found. We call shortcuts that create such separated domains \textit{structural insulators}. In contrast, long shortcuts foster early synchronization of otherwise distant areas of the network, which is why we call them \textit{structural promotors}. 

\begin{figure}
    \centering
    \includegraphics[width=.95\linewidth]{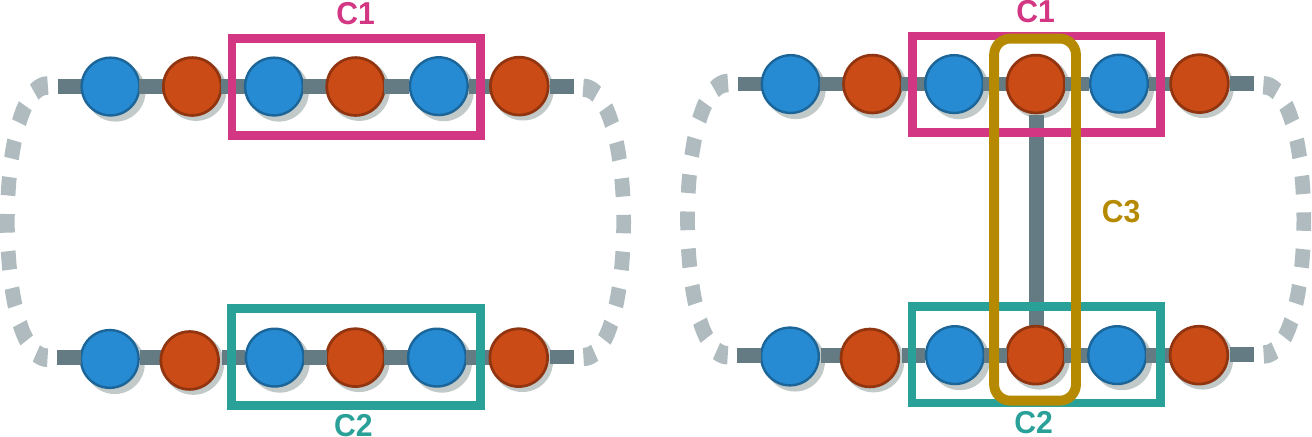}
    \caption{Sketch of a ring graph with two locally correct but globally incompatible coloring domains. (left) Both sides of the ring have a locally correct coloring. For the red node in the first contexture (C1), the only logical color option is to stay red. Likewise, for the red node in the second contexture the only logical option is to stay red. However, since the solutions are globally incompatible, one contexture needs to change their logic in order to reach solved system. (right) Through an inserted link both contextures get the possibility to observe another contexture (another local logic). The color-choice of the connected neighbour is now affected by the own color. In the context of polycontextural logic, the link can hence be interpreted as a third context (C3), that allows self-reflection.}
    \label{fig:pkl}
\end{figure}

The graph coloring problem can also be analyzed as an example of distributed logical systems: The conflicts encountered in graph coloring dynamics on a ring arise due to two (or more) coloring domains that are structurally equal (they are correctly colored) but locally different (they follow a different color permutation).
From a mathematical point of view, this inconsistency between local logical systems relates to distributed logic. Our results can hence be interpreted from the perspective of Gotthard Günther's \textit{theory of polycontexturality} (often also termed \textit{transclassical logic})~\cite{guentherBeitraegeZurGrundlegung1976}.

According to this theory, every interacting subject spans a -- possibly unique -- isolated logic, a \textit{contexture}. All contextures have equal rights and are aligned in a heterarchy. Therefore, no contexture can be said to be right or wrong. In our system, each node can be regarded as a subject (an observer) that spans a contexture. Different logics then show up by the fact that locally correct solutions do not match globally (compare Fig.~\ref{fig:pkl}(left)). However, if the network contains a link as depicted in Fig.~\ref{fig:pkl}(right), then each connected node can observe the respective node of the other contexture: it can observe the results of the observations of another observer. Günther's theory states that these observations of other observers allow for self-reflection and a questioning of one's own logic. In our model, by observing persistent color conflicts with a remote node, nodes gain the ability to recognize that their own color choice does not correspond to the globally valid logic. To select the color randomly instead of based on reasoned considerations can then be understood as a switch of the local logic.

In this view, it also becomes intuitive, why longer shortcuts serve as \textit{promotors} and shorter shortcuts serve as \textit{insulators}: For a node with a shortcut, its ability to self-reflect the own logic requires a link to truly independent information, transcending the local solution regime. 

As a minimal model for the effects of links or information flow within polycontextural systems, the analysis of the graph coloring problem can contribute to heterarchical approaches in biology~\cite{bruniHeterarchicalApproachBiology2015}, consensus finding~\cite{falkCollectivePatternsStable2022}, complex and reflexive relations in social systems~\cite{vogdPolykontexturalitaetErforschungKomplexer2013,jansenANTInfralanguageReflexivity2017}, or transformations in physics~\cite{falkPhysicsOrganizedTransformations2021}.

We also believe that our findings have implications for the
understanding of the emergence of technological standards (here represented by globally
compatible solutions), as well as for the development of
more robust scheduling schemes in manufacturing and
resource distribution~\cite{nandhiniStudyCourseTimetable2019}. 

As emphasized at different points in the text, we focused on a heterarchical model definition. In real systems, however, we often observe hierarchies, where some few nodes have a central position and the ability to coordinate tasks. With respect to graph-coloring dynamics, the effect of such hierarchies was examined in Ref.~\cite{blunckBalanceAutonomousCentralized2018a}. The authors introduce leader nodes, which have a global view of the system and provide an additional source of information for connected nodes. Extending our work to such leader nodes (and to hierarchical networks in general) is an interesting avenue for future work. Indeed, link structures in hierarchical networks may display an overlay of structural insulators and promotors.

\begin{figure}
    \centering
    \includegraphics[width=.95\linewidth]{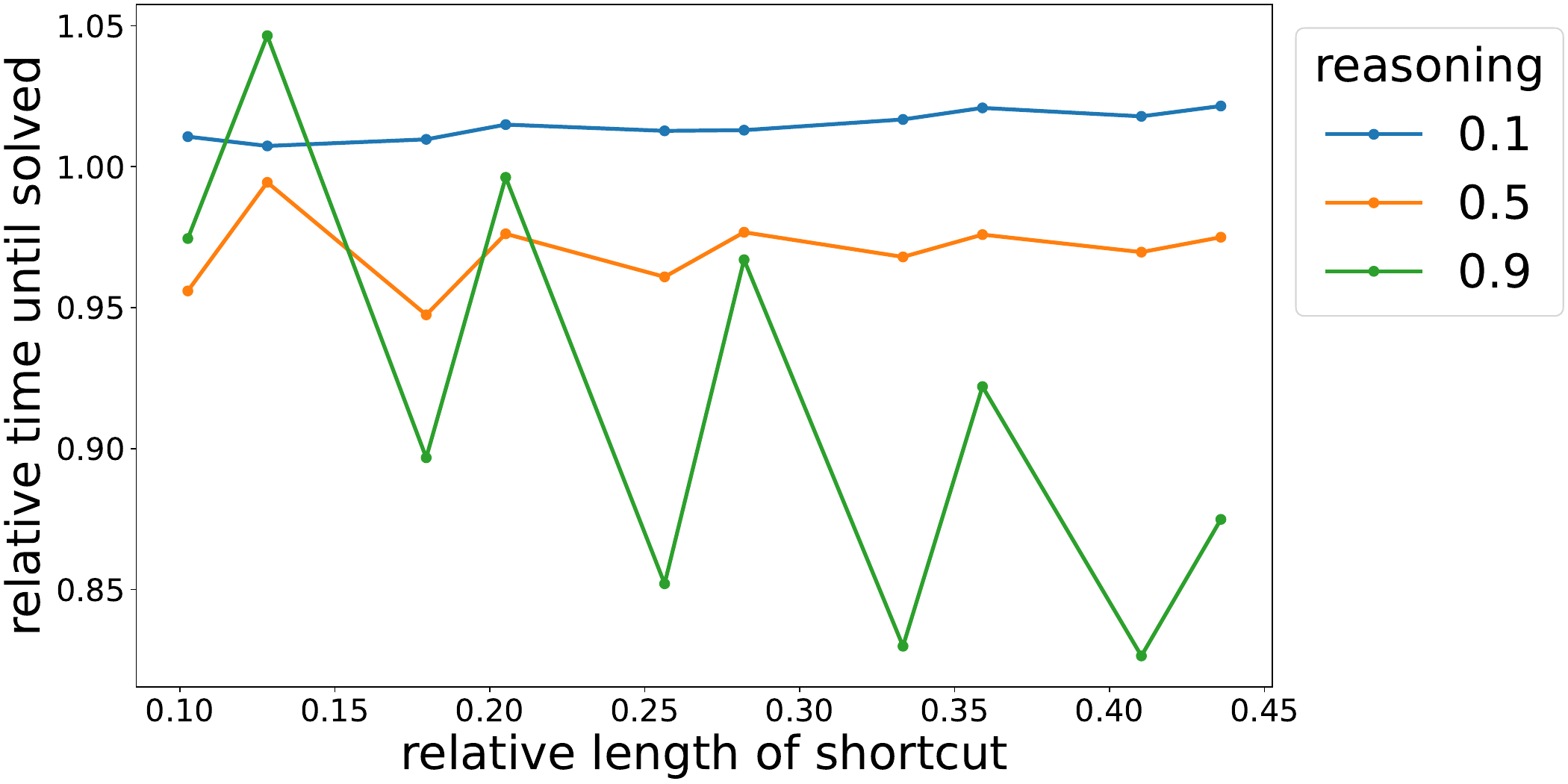}
    \caption{Relative extra time until a ring graph with 39 nodes and a chromatic number of 3 is solved vs. the length of the added shortcut for three different values of $r$. A relative length of $1$ refers to a shortcut of maximal length, hence spanning half of the system. The time is measured relative to the time that is needed if the ring graph does not have any shortcut.}
    \label{fig:chromati3_shortcuts}
\end{figure}

\bibliographystyle{ws-acs}
\bibliography{ws-acs}

\begin{thebibliography}{10}
\providecommand{\urlprefix}{}
\expandafter\ifx\csname urlstyle\endcsname\relax
  \providecommand{\doi}[1]{doi:\discretionary{}{}{}#1}\else
  \providecommand{\doi}{doi:\discretionary{}{}{}\begingroup
  \urlstyle{rm}\Url}\fi

\bibitem{araujoLongrangeConnectionsRealworld2022}
Araújo, T. and Mendes, R.~V., Long-range connections, real-world networks and
  rates of diffusion, \emph{Advances in Complex Systems}  (2022) 2250009.

\bibitem{barabasiEmergenceScalingRandom1999}
Barabási, A.-L. and Albert, R., Emergence of scaling in random networks,
  \emph{Science} \textbf{286} (1999) 509--512.

\bibitem{blunckBalanceAutonomousCentralized2018}
Blunck, H., Armbruster, D., Bendul, J., and H{\"u}tt, M.-T., The balance of
  autonomous and centralized control in scheduling problems, \emph{Applied
  Network Science} \textbf{3} (2018) 16.

\bibitem{bruniHeterarchicalApproachBiology2015}
Bruni, L.~E. and Giorgi, F., Towards a heterarchical approach to biology and
  cognition, \emph{Progress in Biophysics and Molecular Biology} \textbf{119}
  (2015) 481--492.

\bibitem{buttAgentBasedModelingDistributed2020}
Butt, M.~M., Dey, I., Dzaferagic, M., Murphy, M., Kaminski, N., and Marchetti,
  N., Agent-based modeling for distributed decision support in an iot network,
  \emph{IEEE Internet of Things Journal} \textbf{7} (2020) 6919--6931.

\bibitem{caseyDynamicsTwosidedPlatform2012}
Casey, T.~R. and T{\"o}yli, J., Dynamics of two-sided platform success and
  failure: An analysis of public wireless local area access,
  \emph{Technovation} \textbf{32} (2012) 703--716.

\bibitem{falkPhysicsOrganizedTransformations2021}
Falk, J., Eichler, E., Windt, K., and H{\"u}tt, M.-T., Physics is organized
  around transformations connecting contextures in a polycontextural world,
  \emph{Foundations of Science}  (2021).

\bibitem{falkCollectivePatternsStable2022}
Falk, J., Eichler, E., Windt, K., and H{\"u}tt, M.-T., Collective patterns and
  stable misunderstandings in networks striving for consensus without a common
  value system, \emph{Scientific Reports} \textbf{12} (2022) 3028.

\bibitem{grolikInformationLogisticsDecentralized2012}
Grolik, S., \emph{Information Logistics. Decentralized Approaches of
  Information Allocation in Information Exchange Networks} ({ibidem Press},
  Stuttgart, 2012).

\bibitem{guentherBeitraegeZurGrundlegung1976}
G{\"u}nther, G., \emph{Beitr\"age zur Grundlegung einer operationsf\"ahigen
  Dialektik: Metakritik der Logik, nicht-aristotelische Logik, Reflexion,
  Stellenwerttheorie, Dialektik, Cybernetic ontology, Morphogrammatik,
  transklassische Maschinentheorie} ({Felix Meiner Verlag}, 1976).

\bibitem{hadzhievMODELGRAPHCOLORING2009}
Hadzhiev, B., Windt, K., Bergholz, W., and H{\"u}tt, M.-T., A model of graph
  coloring dynamics with attention waves and strategic waiting, \emph{Advances
  in Complex Systems} \textbf{12} (2009) 549--564.

\bibitem{hernandezDistributedGraphColoring2011}
Hern{\'a}ndez, H. and Blum, C., Distributed graph coloring in wireless ad hoc
  networks: A light-weight algorithm based on japanese tree frogs' calling
  behaviour, in \emph{2011 4th Joint IFIP Wireless and Mobile Networking
  Conference (WMNC 2011)} (2011), pp. 1--7, \doi{10.1109/WMNC.2011.6097216}.

\bibitem{jansenANTInfralanguageReflexivity2017}
Jansen, T., Beyond ant: Towards an `infra-language' of reflexivity,
  \emph{European Journal of Social Theory} \textbf{20} (2017) 199--215.

\bibitem{juddBehavioralDynamicsInfluence2010b}
Judd, S., Kearns, M., and Vorobeychik, Y., Behavioral dynamics and influence in
  networked coloring and consensus, \emph{Proceedings of the National Academy
  of Sciences} \textbf{107} (2010) 14978--14982.

\bibitem{kearnsExperimentalStudyColoring2006}
Kearns, M., An experimental study of the coloring problem on human subject
  networks, \emph{Science} \textbf{313} (2006) 824--827.

\bibitem{kearnsBehavioralExperimentsBiased2009b}
Kearns, M., Judd, S., Tan, J., and Wortman, J., Behavioral experiments on
  biased voting in networks, \emph{Proceedings of the National Academy of
  Sciences} \textbf{106} (2009) 1347--1352.

\bibitem{leithSelfManagedDistributedChannel2006a}
Leith, D. and Clifford, P., A self-managed distributed channel selection
  algorithm for wlans, in \emph{2006 4th International Symposium on Modeling
  and Optimization in Mobile, Ad Hoc and Wireless Networks} (2006), pp. 1--9,
  \doi{10.1109/WIOPT.2006.1666484}.

\bibitem{marrSimilarImpactTopological2006}
Marr, C. and H{\"u}tt, M.-T., Similar impact of topological and dynamic noise
  on complex patterns, \emph{Physics Letters A} \textbf{349} (2006) 302--305.

\bibitem{nandhiniStudyCourseTimetable2019}
Nandhini, V., A study on course timetable scheduling and exam timetable
  scheduling using graph coloring approach, \emph{International Journal for
  Research in Applied Science and Engineering Technology} \textbf{7} (2019)
  1999--2006.

\bibitem{papachristosSystemDynamicsModel2021}
Papachristos, G. and {van de Kaa}, G., A system dynamics model of standards
  competition, \emph{IEEE Transactions on Engineering Management} \textbf{68}
  (2021) 18--32.

\bibitem{shiradoLocallyNoisyAutonomous2017}
Shirado, H. and Christakis, N.~A., Locally noisy autonomous agents improve
  global human coordination in network experiments, \emph{Nature} \textbf{545}
  (2017) 370--374.

\bibitem{suarezDominantDesignsSurvival1995}
Su{\'a}rez, F.~F. and Utterback, J.~M., Dominant designs and the survival of
  firms, \emph{Strategic Management Journal} \textbf{16} (1995) 415--430.

\bibitem{vandekaaFactorsWinningFormat2015}
{van de Kaa}, G. and {de Vries}, H.~J., Factors for winning format battles: A
  comparative case study, \emph{Technological Forecasting and Social Change}
  \textbf{91} (2015) 222--235.

\bibitem{vandekaaEmergenceStandardsMetaAnalysis2007}
{van de Kaa}, G., De~Vries, H.~J., {van Heck}, E., and {van den Ende}, J., The
  emergence of standards: A meta-analysis, in \emph{2007 40th Annual Hawaii
  International Conference on System Sciences (HICSS'07)} ({IEEE}, {Waikoloa,
  HI}, 2007), ISBN 978-0-7695-2755-0, pp. 173a--173a,
  \doi{10.1109/HICSS.2007.529}.

\bibitem{vogdPolykontexturalitaetErforschungKomplexer2013}
Vogd, W., Polykontexturalit\"at: Die erforschung komplexer systemischer
  zusammenh\"ange in theorie und praxis, \emph{Familiendynamik} \textbf{38}
  (2013) 32--41.

\bibitem{wattsCollectiveDynamicsSmallworld1998}
Watts, D.~J. and Strogatz, S.~H., Collective dynamics of `small-world'
  networks, \emph{Nature} \textbf{393} (1998) 440--442.

\bibitem{windtGraphColoringDynamics2010}
Windt, K. and H{\"u}tt, M.~T., Graph coloring dynamics: A simple model scenario
  for distributed decisions in production logistics, \emph{CIRP Annals}
  \textbf{59} (2010) 461--464.

\end{thebibliography}

\end{document}